\documentclass[conference]{IEEEtran}
\IEEEoverridecommandlockouts
\usepackage{cite}
\usepackage{amsmath,amssymb,amsfonts}
\usepackage{algorithmic}
\usepackage{graphicx}
\usepackage{textcomp}
\usepackage{xcolor}
\usepackage{url}
\usepackage{hyperref}
\def\BibTeX{{\rm B\kern-.05em{\sc i\kern-.025em b}\kern-.08em
    T\kern-.1667em\lower.7ex\hbox{E}\kern-.125emX}}
\begin{document}

\title{Continuous surrogates versus threshold Boolean networks for modeling Arabidopsis ISR gene regulation\\

\thanks{The author thanks ANID FONDECYT 1230315, ANID-MILENIO-NCN2024\_103, ANID-MILENIO-NCN2024\_047, Centro de Modelamiento Matemático (CMM) FB210005, BASAL funds for centers of excellence from ANID-Chile, and the Data Observatory Foundation (DO).}
}

\author{\IEEEauthorblockN{Gonzalo A. Ruz}
\IEEEauthorblockA{\textit{Facultad de Ingeniería y Ciencias, Universidad Adolfo Ib\'{a}\~{n}ez} \\
\textit{Millennium Nucleus for Social Data Science (SODAS)}\\
\textit{Millennium Nucleus in Data Science for Plant Resilience (PhytoLearning)}\\
Santiago, Chile \\
gonzalo.ruz@uai.cl}

}

\maketitle

\begin{abstract}
Gene regulatory network modeling often requires balancing predictive accuracy and mechanistic interpretability. In this work, we compare continuous surrogate models and a discrete mechanistic model on the same \textit{Arabidopsis thaliana} induced systemic resistance (ISR) dataset, using both the raw continuous gene-expression measurements and their sign-binarized representation. The study considers eight defense-related genes measured over nine time points and evaluates two continuous predictors, Random Forest (RF) regression and a Multi-Layer Perceptron (MLP), against a threshold Boolean network (TBN). The models are assessed using rolling-origin one-step prediction, recursive multi-step rollout, and interpretability analysis. RF achieved the best average one-step numerical performance in the continuous domain, with an MAE of 1.910 and an RMSE of 2.836, compared with 2.089 and 3.106 for the MLP. In the binary domain, the TBN obtained the best average one-step qualitative performance, with a binary accuracy of 0.550 and a Hamming distance of 3.600, compared with 0.500 and 4.000 for RF, and 0.495 and 4.040 for the MLP. In recursive rollout, the TBN exactly reproduced the observed binarized trajectory, while the MLP also showed near-perfect fidelity, with a trajectory binary accuracy of 0.986, and RF accumulated substantially larger deviation, with a trajectory binary accuracy of 0.708. These results highlight that local numerical accuracy and global qualitative dynamical fidelity are not necessarily aligned, and suggest that continuous surrogates and threshold Boolean networks should be viewed as complementary tools for modeling biological regulation.
\end{abstract}


\section{Introduction}

Gene regulatory networks (GRNs) provide a systems-level view of how genes coordinate cellular responses to developmental, environmental, and stress signals. Because GRNs are inherently dynamic, their study requires models that can represent not only pairwise associations, but also temporal state transitions and stable regulatory configurations. In recent years, machine learning has considerably expanded the methodological landscape of GRN inference, especially with the growth of transcriptomic and single-cell data. A recent review categorizes modern GRN inference methods across supervised, unsupervised, semi-supervised, and contrastive-learning paradigms, and highlights the rapid rise of deep learning approaches for extracting regulatory structure from increasingly rich omics datasets \cite{Hegde2025MLGRNReview}. At the same time, this progress has sharpened a long-standing tension in computational biology: more flexible predictive models often improve local predictive performance, but they may be harder to interpret mechanistically.

Boolean and threshold-based models \cite{6217257} remain attractive in this context because they offer an explicit dynamical representation of gene regulation while retaining interpretability. In Boolean networks, genes are represented as active/inactive variables and the system evolves through discrete update rules, making attractors and trajectories directly analyzable. This modeling paradigm continues to be actively used in contemporary applications. Recent examples include Boolean network models for breast cancer therapeutic target optimization \cite{Sgariglia2024BreastCancerBN}, Boolean models of human preimplantation development built from single-cell transcriptomic data \cite{Bolteau2024PreimplantationBN}, and recent data-driven methodologies that combine transcriptomic evidence and prior knowledge to infer Boolean models capable of reproducing differentiation and reprogramming behavior \cite{Chevalier2025BNTranscriptomes}. In plant systems, Boolean modeling has also been recognized as a useful framework for representing developmental and signaling processes when mechanistic detail is incomplete but qualitative regulation is known \cite{Karanam2022BooleanPlants}.

Among Boolean-model variants, threshold Boolean networks (TBNs) \cite{RUZ2025105572,RUZ2026105682} are especially appealing because they replace potentially complex logical rules with weighted sums and thresholds. This makes them easier to train from data and often easier to interpret biologically. However, recent work has also made clear that this simplification comes with limitations. In particular, a recent systematic assessment of threshold formalisms showed that simple threshold rules can approximate the dynamics of some expert-curated biological GRNs, especially smaller and less connected ones, but they may fail to fully recover the logic and dynamics of larger or highly connected networks \cite{Kadelka2025ThresholdAssessment}. This makes TBNs an interesting compromise: they are mechanistically structured and interpretable, yet their expressive power is not unlimited. Understanding when such models are sufficient, and how they compare with more flexible continuous predictors, remains an open and practically important question.

This tradeoff is particularly relevant in settings where both continuous expression measurements and biologically meaningful binary abstractions are available. One such case is the induced systemic resistance (ISR) response of \textit{Arabidopsis thaliana} triggered by the beneficial bacterium \textit{Paraburkholderia phytofirmans} PsJN. ISR is a cost-effective plant defense strategy in which beneficial microbes prime the plant for enhanced resistance against future pathogen attack. In the Arabidopsis--PsJN system, previous work reconstructed a GRN for the ISR defense response using time-series expression data from eight defense-related genes and a threshold Boolean framework, showing that discrete dynamical modeling could recover plausible interactions and biologically meaningful trajectories \cite{Timmermann2020ISR}. More recent transcriptomic work has continued to show that \textit{P. phytofirmans} PsJN induces broad local and systemic transcriptional reprogramming in \textit{Arabidopsis}, reinforcing the biological relevance of this host--microbe interaction and the need for models that can connect expression patterns to regulatory mechanisms \cite{Nesic2025PsJN}.

At the same time, the broader GRN-inference literature increasingly emphasizes flexible machine learning models that can adapt to nonlinear and context-specific regulation. For example, recent approaches such as TIGER model activation and inhibition in a flexible Bayesian framework to improve context-specific transcription-factor activity estimation \cite{Chen2024TIGER}, and methods such as LINGER use neural networks to infer GRNs from paired single-cell expression and chromatin accessibility data \cite{Yuan2025LINGER}. Beyond GRN inference per se, the regulatory-genomics community has also begun to use surrogate-model ideas explicitly to obtain interpretable approximations of complex predictors, including domain-specific surrogate modeling for genomic deep networks \cite{Seitz2024SQUID}. These developments suggest a promising but still underexplored direction: using continuous surrogate models not as replacements for mechanistic GRN models, but as complementary representations whose predictive behavior can be contrasted with discrete regulatory dynamics.

In this work, we explore that comparison directly. Using the same Arabidopsis ISR dataset in two forms, its raw continuous expression measurements and a sign-binarized representation, we compare two continuous surrogate models, Random Forest (RF) regression and a Multi-Layer Perceptron (MLP), with a discrete mechanistic model based on a threshold Boolean network. Our goal is not simply to rank predictors by error, but to study the tradeoff between local numerical prediction, qualitative trajectory fidelity, and interpretability. In particular, we ask whether models that perform well on one-step prediction in the continuous domain also preserve the biologically meaningful qualitative dynamics of the ISR response, and how this behavior compares with that of an explicitly mechanistic threshold Boolean representation.

The main contributions of this paper are threefold. First, we formulate a unified comparison between continuous and discrete models using the same underlying biological time series in two representations. Second, we evaluate the models not only through one-step prediction metrics, but also through recursive rollout fidelity and qualitative agreement with the observed binary trajectory. Third, we provide an interpretable comparison between surrogate-based and threshold-based views of the same regulatory process, thereby clarifying the extent to which predictive flexibility and mechanistic transparency align or diverge in this case study.

The rest of the paper is organized as follows. Section II presents the Methods, including the dataset representation, the continuous surrogate models, the threshold Boolean network, and the evaluation protocol. Section III reports the experimental Results, covering one-step prediction performance, multi-step rollout fidelity, and interpretability analyses. Section IV discusses the main findings, limitations, and implications of the comparison between continuous and discrete modeling approaches. Finally, Section V concludes the paper and outlines directions for future work.

\section{Methods}\label{meds}

\subsection{Dataset and problem formulation}

We considered the \textit{Arabidopsis thaliana} induced systemic resistance (ISR) dataset composed of expression measurements for eight defense-related genes:
\textit{PR1}, \textit{PDF1.2}, \textit{WRKY70}, \textit{WRKY54}, \textit{WRKY33}, \textit{MYC2}, \textit{ERF1}, and \textit{LOX2}.
The data consist of nine temporal observations collected at
$t \in \{0, 0.5, 1, 3, 6, 9, 12, 18, 24\}$ hours, yielding a multivariate time series with eight variables and nine time points.

Let $\mathbf{x}(t) \in \mathbb{R}^8$ denote the continuous gene-expression vector at time $t$.
From the observed sequence, we constructed eight one-step transitions of the form
\[
\mathbf{x}(t_k) \rightarrow \mathbf{x}(t_{k+1}), \qquad k=1,\dots,8.
\]
These transition pairs were used to define the prediction task for the continuous models.
In parallel, a binarized version of the same dataset was constructed to train and evaluate the threshold Boolean network (TBN), allowing both continuous and discrete modeling paradigms to be compared on the same biological process.

The main objective of the study is to compare continuous machine-learning surrogates and a discrete mechanistic model in terms of:
(i) one-step predictive performance,
(ii) reproduction of the observed temporal trajectory,
(iii) qualitative dynamical behavior under iterative rollout, and
(iv) interpretability of the inferred regulatory structure.

\subsection{Continuous and discrete data representations}

Two representations of the same dataset were considered.

\paragraph{Continuous representation.}
The original real-valued expression measurements were used to train continuous surrogate models.
For each transition, the input is the expression vector at time $t_k$ and the target is the expression vector at time $t_{k+1}$.

\paragraph{Discrete representation.}
A binarized version of the dataset was obtained from the continuous measurements by applying a sign-based discretization rule:
\[
x_i^{(b)}(t)=
\begin{cases}
1, & \text{if } x_i(t) > 0,\\
0, & \text{otherwise},
\end{cases}
\]
for gene $i$ at time $t$.
This binary representation was used to train and evaluate the TBN.

Because the sampling times are not equally spaced, we included the time increment
\[
\Delta t_k = t_{k+1} - t_k
\]
as an additional input variable for the continuous models.
Thus, for transition $k$, the continuous predictor input is
\[
\tilde{\mathbf{x}}(t_k) = [\mathbf{x}(t_k), \Delta t_k] \in \mathbb{R}^9,
\]
whereas the TBN operates only on the binarized gene states and does not explicitly use $\Delta t_k$.

\subsection{Continuous surrogate models}

Two continuous surrogate models were considered: a Random Forest regressor and a Multi-Layer Perceptron (MLP) regressor.
Both models were trained in a multi-output setting to predict the full eight-dimensional expression vector at the next time point.

\subsubsection{Random Forest regression}

Let $f_{\mathrm{RF}}$ denote the Random Forest predictor. Given the current expression vector and the corresponding time increment, the model predicts the next continuous state as
\[
\hat{\mathbf{x}}(t_{k+1}) = f_{\mathrm{RF}}(\tilde{\mathbf{x}}(t_k)).
\]
The Random Forest was implemented using 300 trees, maximum depth equal to 3, and random state 0. Random Forests were selected because they can capture nonlinear relationships, are relatively robust in small-sample settings, and do not require feature scaling.

\subsubsection{Multi-Layer Perceptron regression}

Let $f_{\mathrm{MLP}}$ denote the neural surrogate. The MLP receives the same input vector $\tilde{\mathbf{x}}(t_k)$ and predicts the next continuous expression state,
\[
\hat{\mathbf{x}}(t_{k+1}) = f_{\mathrm{MLP}}(\tilde{\mathbf{x}}(t_k)).
\]
The MLP used a single hidden layer with 12 units, ReLU activation, L2 regularization parameter $\alpha = 10^{-2}$, the \texttt{lbfgs} optimizer, and a maximum of 5000 iterations.

Because neural networks are sensitive to feature scale, both inputs and targets were standardized using \texttt{StandardScaler}, with scaling parameters estimated only from the training portion of each evaluation split.
To account for sensitivity to initialization, the MLP was trained repeatedly with random seeds $\{0,1,2,3,4\}$ during the rolling-origin evaluation.

\subsection{Threshold Boolean network model}

To model the same biological process in a discrete mechanistic way, we used a threshold Boolean network (TBN) defined over the binarized gene states.
Let $\mathbf{x}^{(b)}(t) \in \{0,1\}^8$ denote the binary state of the network at time $t$.
The state of node $i$ at the next time step is updated according to
\[
x_i^{(b)}(t+1) =
\begin{cases}
1, & \text{if } \sum_{j=1}^{8} w_{ji} x_j^{(b)}(t) + b_i \ge 0,\\
0, & \text{otherwise},
\end{cases}
\]
where $w_{ji}$ is the influence of node $j$ on node $i$, and $b_i$ is the bias term of node $i$.

In vector form, the update rule is
\[
\mathbf{x}^{(b)}(t+1) = H\!\left(\mathbf{x}^{(b)}(t)W + \mathbf{b}\right),
\]
where $W \in \mathbb{R}^{8 \times 8}$ is the weight matrix, $\mathbf{b} \in \mathbb{R}^8$ is the bias vector, and $H(\cdot)$ is the Heaviside function applied element-wise.

The TBN was trained on the binarized transition pairs
\[
\mathbf{x}^{(b)}(t_k) \rightarrow \mathbf{x}^{(b)}(t_{k+1}),
\]
using a node-wise perceptron strategy \cite{10944722}. Specifically, one binary linear-threshold classifier was trained independently for each target gene, using as input the current binarized 8-gene state and as target the next binarized value of that gene. The learned coefficient vector for each gene forms one column of $W$, and the corresponding learned bias forms one component of $\mathbf{b}$.

To avoid complications associated with one-class training folds in such a small dataset, a manual perceptron implementation was used. For each target gene, the parameters were initialized at zero and updated according to the classical perceptron rule. Given a training pair $(\mathbf{x}, y)$ with $\mathbf{x}\in\{0,1\}^8$ and $y\in\{0,1\}$, the prediction is
\[
\hat{y} =
\begin{cases}
1, & \text{if } \mathbf{x}^{\top}\mathbf{w} + b \ge 0,\\
0, & \text{otherwise},
\end{cases}
\]
and the update is
\[
\mathbf{w} \leftarrow \mathbf{w} + \eta (y-\hat{y})\mathbf{x}, \qquad
b \leftarrow b + \eta (y-\hat{y}),
\]
with learning rate $\eta=1$.
Training was run for at most 100 epochs and stopped early if no classification errors occurred in an epoch.
When a target gene had only one class in the current training fold, the corresponding node was modeled by a constant binary predictor.

This perceptron-based node-wise training procedure is in the same spirit as earlier threshold Boolean network training approaches based on one classifier per gene, followed by assembly of the global network parameters \cite{10944722}.

\subsection{Evaluation protocol}

Given the very small size of the dataset, we did not use random train/test partitions.
Instead, we adopted a rolling-origin evaluation scheme based on the temporal ordering of the data.

Specifically, for each evaluation step, models were trained on an expanding set of initial transitions and tested on the immediately following unseen transition.
The minimum training size was set to 3 transitions, yielding test cases for the transitions
$3\rightarrow6$, $6\rightarrow9$, $9\rightarrow12$, $12\rightarrow18$, and $18\rightarrow24$ hours.
This protocol preserves temporal causality and is more appropriate for short time-series forecasting than random resampling.

For each split:
\begin{itemize}
    \item RF was trained once using random state 0;
    \item MLP was trained five times using random seeds $\{0,1,2,3,4\}$;
    \item TBN was trained once on the corresponding binarized training transitions.
\end{itemize}

The reported MLP one-step results aggregate the performance obtained across these five seeds.
RF and TBN were evaluated once per split.

\subsection{One-step prediction metrics}

The continuous models were first evaluated on the raw expression prediction task.
Given the true next state $\mathbf{x}(t_{k+1})$ and the predicted state $\hat{\mathbf{x}}(t_{k+1})$, we computed the mean absolute error (MAE) and root mean squared error (RMSE):
\[
\mathrm{MAE} = \frac{1}{8}\sum_{i=1}^{8} \left|x_i(t_{k+1}) - \hat{x}_i(t_{k+1})\right|,
\]
\[
\mathrm{RMSE} = \sqrt{\frac{1}{8}\sum_{i=1}^{8} \left(x_i(t_{k+1}) - \hat{x}_i(t_{k+1})\right)^2 }.
\]

To compare the continuous surrogates with the TBN on a common qualitative basis, the predictions of RF and MLP were binarized by sign:
\[
\hat{x}_i^{(b)}(t_{k+1}) =
\begin{cases}
1, & \text{if } \hat{x}_i(t_{k+1}) > 0,\\
0, & \text{otherwise}.
\end{cases}
\]
These binarized predictions were compared against the observed binary next state using binary accuracy and Hamming distance:
\[
\mathrm{Acc}_{\mathrm{bin}} = \frac{1}{8}\sum_{i=1}^{8} \mathbb{I}\!\left(\hat{x}_i^{(b)}(t_{k+1}) = x_i^{(b)}(t_{k+1})\right),
\]
\[
\mathrm{Ham} = \sum_{i=1}^{8} \mathbb{I}\!\left(\hat{x}_i^{(b)}(t_{k+1}) \neq x_i^{(b)}(t_{k+1})\right),
\]
where $\mathbb{I}(\cdot)$ is the indicator function.

The TBN was evaluated directly on the binary one-step prediction task using the same binary accuracy and Hamming distance metrics.

\subsection{Multi-step rollout analysis}

To assess whether good one-step predictions translate into faithful dynamical behavior, we performed recursive multi-step rollout experiments.

After the rolling-origin evaluation, final models were refit using all eight available transitions.
The final RF was trained once on the full continuous dataset.
The final TBN was trained once on the full binarized dataset.
For the MLP, five final models were trained using seeds $\{0,1,2,3,4\}$ on the full continuous dataset.

Starting from the first observed state, each model was iteratively applied to generate future states.
For the continuous models,
\[
\hat{\mathbf{x}}(t_{k+1}) = f(\hat{\mathbf{x}}(t_k), \Delta t_k),
\]
where the predicted state at one step is fed back as input to the next step together with the corresponding observed time increment.
For the MLP, the rollout used the mean trajectory obtained by averaging the five seed-specific rollout trajectories pointwise.
For the TBN,
\[
\hat{\mathbf{x}}^{(b)}(t_{k+1}) = H\!\left(\hat{\mathbf{x}}^{(b)}(t_k)W + \mathbf{b}\right),
\]
and the model was iterated for the same number of steps as in the observed sequence.

The rollout analysis was used to examine:
(i) reproduction of the observed temporal sequence,
(ii) accumulation of prediction errors over time,
(iii) convergence, oscillation, or drift under repeated iteration, and
(iv) agreement with the qualitative biological progression described by the data.

To compare all models in a common domain, the RF and MLP rollout trajectories were binarized by sign and compared with the observed binarized sequence and the TBN rollout.

\subsection{Interpretability analysis}

Interpretability was examined for the Random Forest and the TBN.

For the Random Forest, feature importance scores were extracted from the fitted model. Since the implementation yields a single importance vector for the multi-output regressor, that vector was replicated across outputs to form a feature-importance matrix for visualization. The input variables considered in this analysis were the eight genes plus $\Delta t$.

For the TBN, interpretability is intrinsic: the sign and magnitude of the learned weights encode putative activating or inhibitory interactions, while the bias terms define the decision boundaries governing state transitions. The learned weight matrix was visualized as a signed interaction map.

This analysis allows the comparison to go beyond predictive performance and address whether the inferred models yield biologically meaningful regulatory hypotheses.

\subsection{Study goal}

The purpose of this study is not to establish a universal ranking among modeling approaches, but rather to characterize the tradeoff between predictive fidelity and mechanistic interpretability in a small but biologically meaningful gene-expression time series.
Continuous surrogate models are expected to better capture expression magnitudes, whereas the TBN is expected to provide a more explicit qualitative representation of regulatory dynamics.

\section{Results}

We compared two continuous surrogate models, namely Random Forest (RF) regression and a Multi-Layer Perceptron (MLP), against a discrete mechanistic model based on a threshold Boolean network (TBN). The comparison was carried out on the Arabidopsis ISR dataset using both the continuous expression profiles and their binarized counterpart. The ISR dataset contains eight defense-related genes measured at nine time points, yielding eight one-step transitions for evaluation. The binarized representation follows the sign-based discretization used in the original ISR study. \cite{Timmermann2020ISR}

\subsection{One-step rolling-origin prediction}

Table~\ref{tab:one_step_summary} summarizes the one-step rolling-origin results across all evaluation splits. For the continuous prediction task, RF achieved the best overall numerical performance, with lower mean MAE and RMSE than the MLP. In particular, RF obtained an average MAE of 1.910 and an average RMSE of 2.836, while the MLP obtained an average MAE of 2.089 and an average RMSE of 3.106.

When the predicted continuous outputs of RF and MLP were binarized by sign and compared against the observed binary next state, both models achieved similar average binary accuracies, with RF reaching 0.500 and MLP 0.495. The TBN yielded the best one-step binary performance, with an average binary accuracy of 0.550 and the lowest mean Hamming distance (3.60). Thus, although the continuous models were able to predict raw expression values, the discrete TBN provided the best qualitative one-step agreement with the observed binarized transitions.

\begin{table}[t]
\centering
\caption{Overall one-step rolling-origin results.}
\label{tab:one_step_summary}
\setlength{\tabcolsep}{3pt}
\footnotesize
\begin{tabular}{lcccc}
\hline
Model & MAE & RMSE & Bin. Acc. & Hamm. Dist. \\
\hline
MLP & $2.089 \pm 0.942$ & $3.106 \pm 1.321$ & $0.495 \pm 0.189$ & $4.040 \pm 1.513$ \\
RF  & $1.910 \pm 0.886$ & $2.836 \pm 1.147$ & $0.500 \pm 0.234$ & $4.000 \pm 1.871$ \\
TBN & -- & -- & $0.550 \pm 0.112$ & $3.600 \pm 0.894$ \\
\hline
\end{tabular}
\end{table}

\begin{figure*}[t]
    \centering
    \includegraphics[scale=0.35]{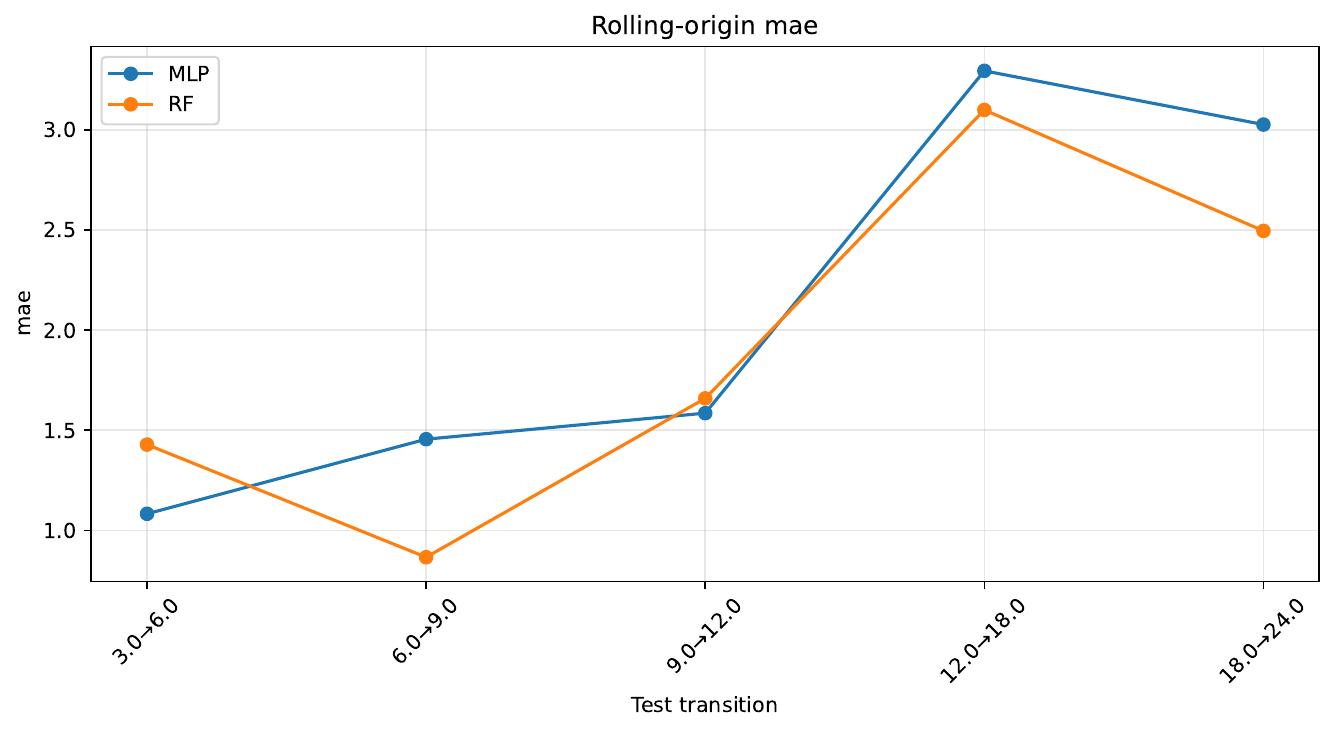}
    \includegraphics[scale=0.35]{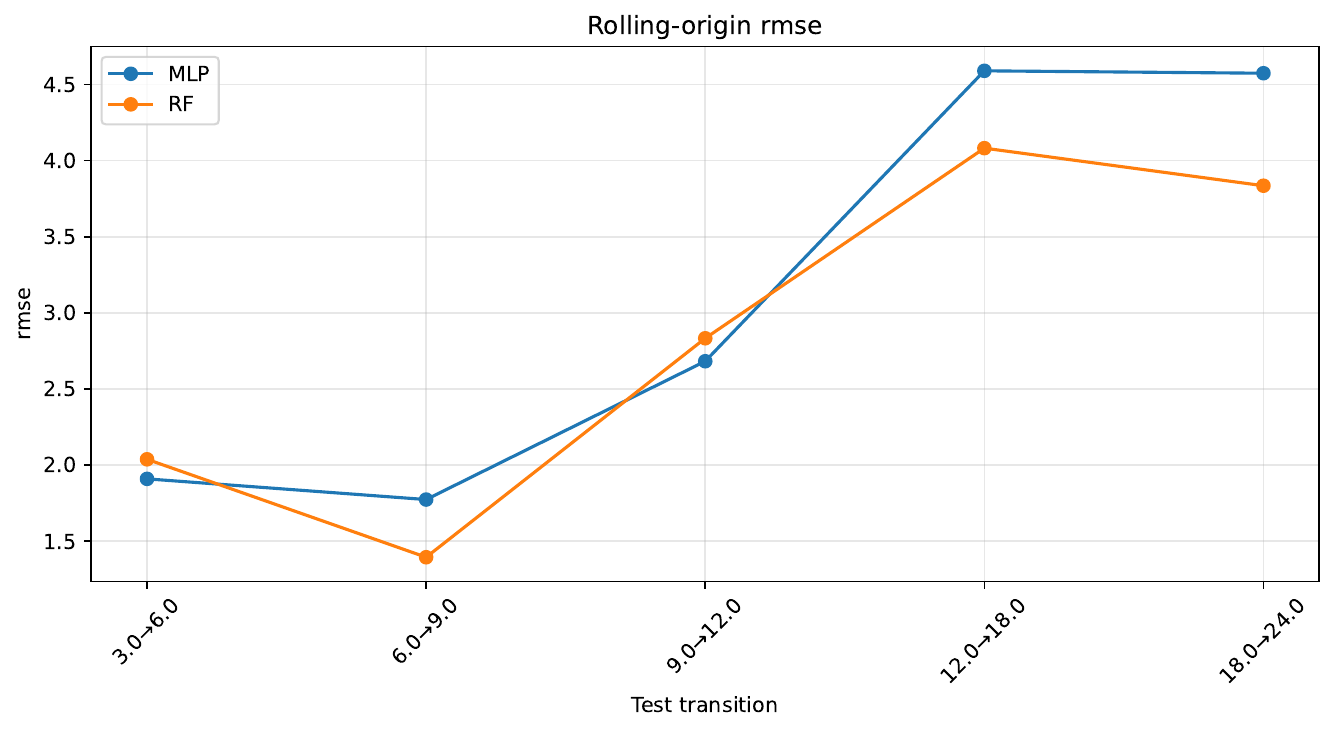}
    \includegraphics[scale=0.35]{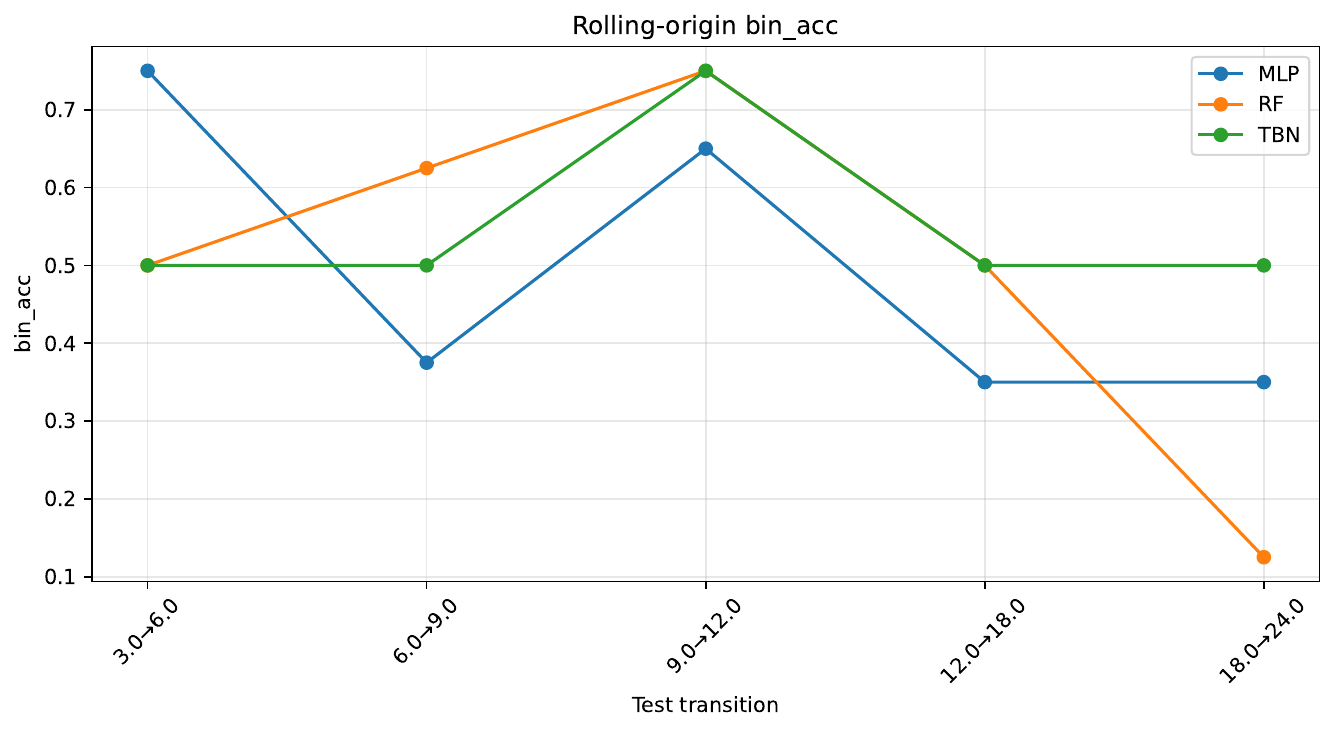}
    \includegraphics[scale=0.35]{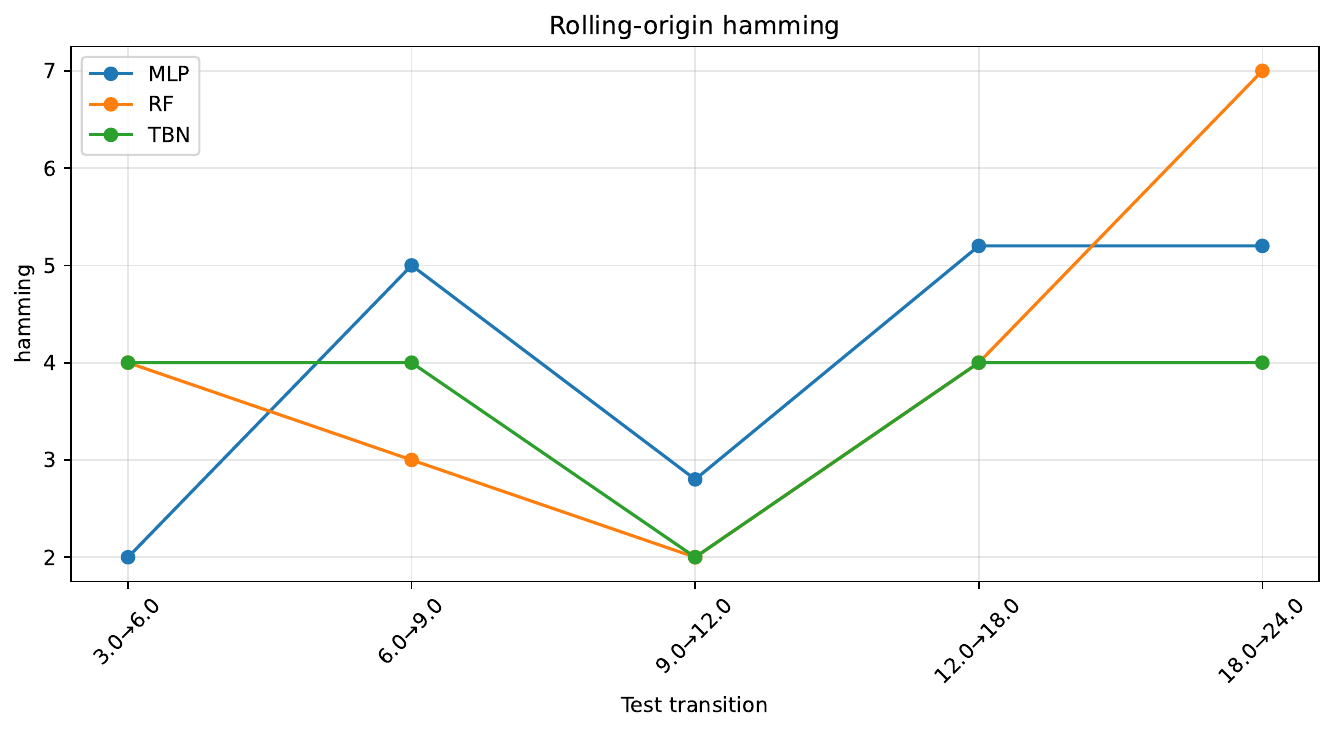}
    \caption{Rolling-origin evaluation across test transitions. Top: MAE and RMSE. Bottom: binary accuracy and Hamming distance. RF and MLP are evaluated in the continuous domain and then binarized for the qualitative comparison, while the TBN is evaluated directly in the binary domain.}
    \label{fig:rolling_origin_metrics}
\end{figure*}

Figure~\ref{fig:rolling_origin_metrics} shows the rolling-origin performance by test transition. The per-split curves reveal that performance was not uniform across time. For example, for the transition from 3~h to 6~h, the MLP outperformed RF in both MAE and RMSE, while for later transitions RF tended to be more stable numerically. In the binary evaluation, the TBN was generally competitive and often superior in terms of Hamming distance.

\subsection{Multi-step rollout fidelity}

To assess whether one-step predictive performance translated into faithful long-range dynamics, each model was recursively rolled out starting from the first observed state. Table~\ref{tab:rollout_fidelity} reports the resulting trajectory fidelity in the binary domain.

The TBN reproduced the observed binarized trajectory exactly, achieving a trajectory binary accuracy of 1.000 and zero Hamming error. The MLP was also highly accurate in rollout, reaching a trajectory binary accuracy of 0.986 with only one binary mismatch across the entire trajectory. In contrast, RF accumulated substantially larger rollout error, with a trajectory binary accuracy of 0.708 and a total Hamming distance of 21.

These results indicate that one-step numerical accuracy alone did not fully determine long-range qualitative fidelity. Although RF had the best average MAE and RMSE in one-step prediction, its recursive rollout diverged more strongly from the observed biological trajectory than both the MLP and the TBN.

\begin{table}[t]
\centering
\caption{Binary-domain multi-step rollout fidelity.}
\label{tab:rollout_fidelity}
\setlength{\tabcolsep}{3pt}
\footnotesize
\begin{tabular}{lccc}
\hline
Model & Traj. Acc. & Tot. Hamm. & Mean Hamm./Time \\
\hline
RF rollout  & 0.7083 & 21 & 2.3333 \\
MLP rollout & 0.9861 & 1  & 0.1111 \\
TBN rollout & 1.0000 & 0  & 0.0000 \\
\hline
\end{tabular}
\end{table}

\begin{figure*}[t]
    \centering
    \includegraphics[scale=0.35]{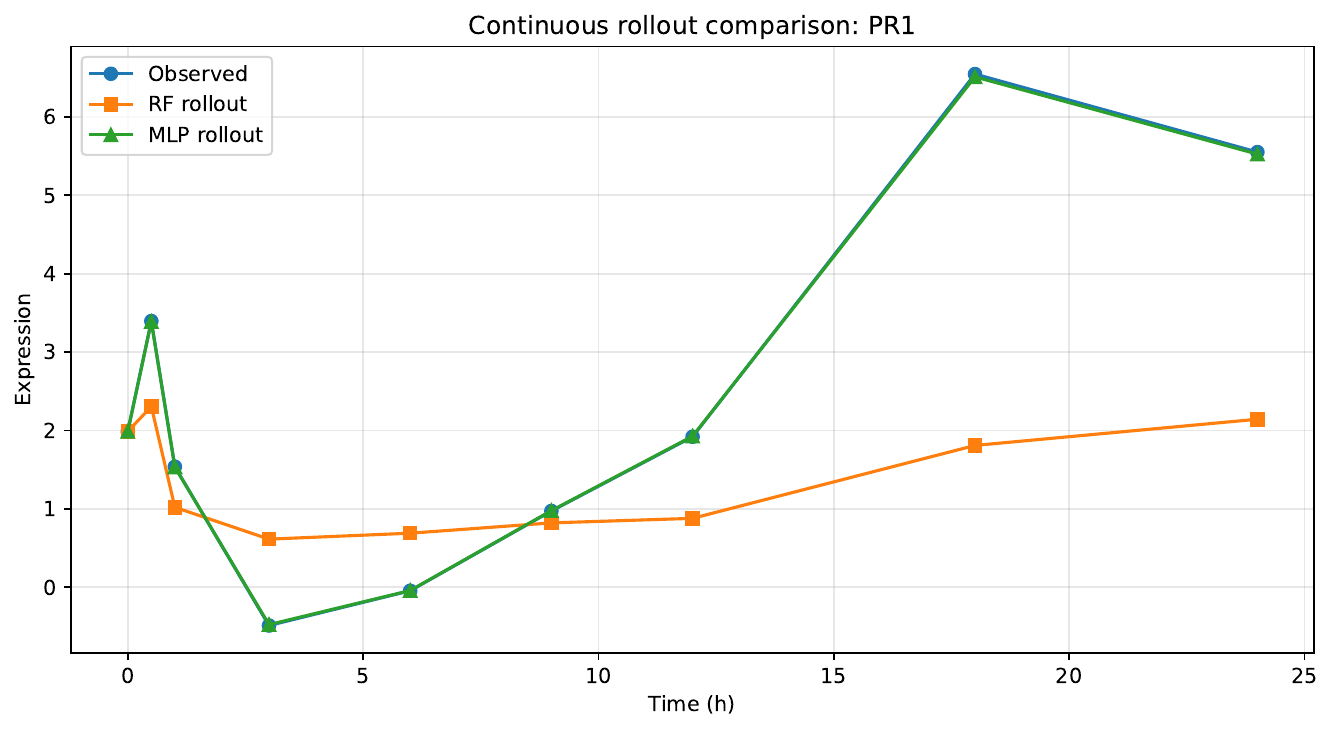}
    \includegraphics[scale=0.35]{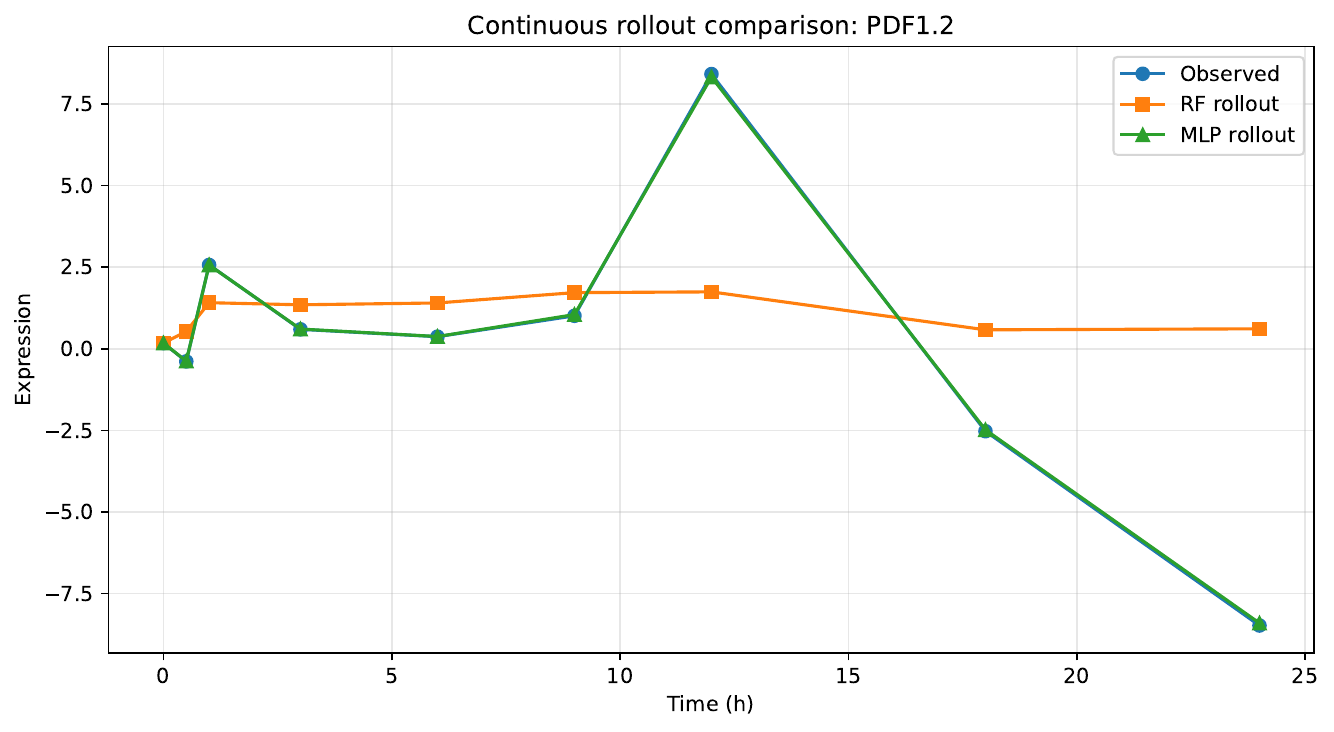}
    \includegraphics[scale=0.35]{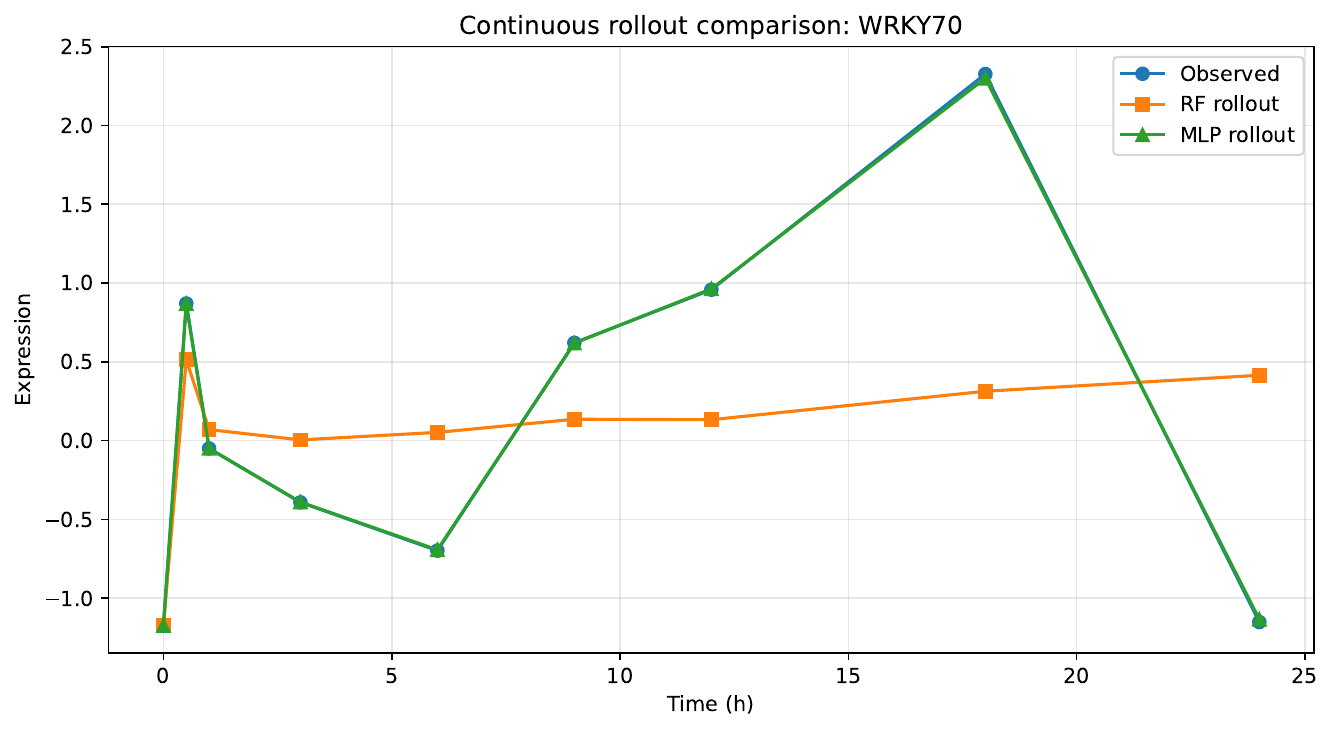}
    \includegraphics[scale=0.35]{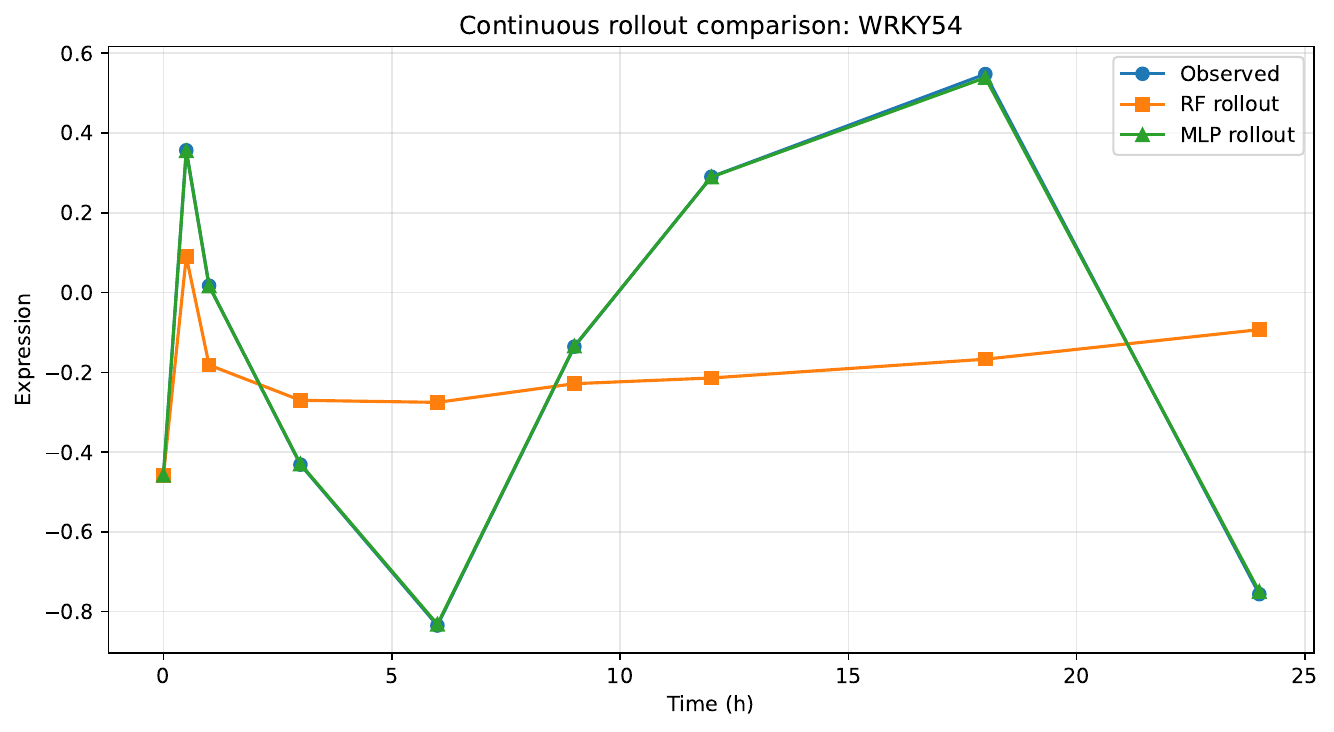}
    \includegraphics[scale=0.35]{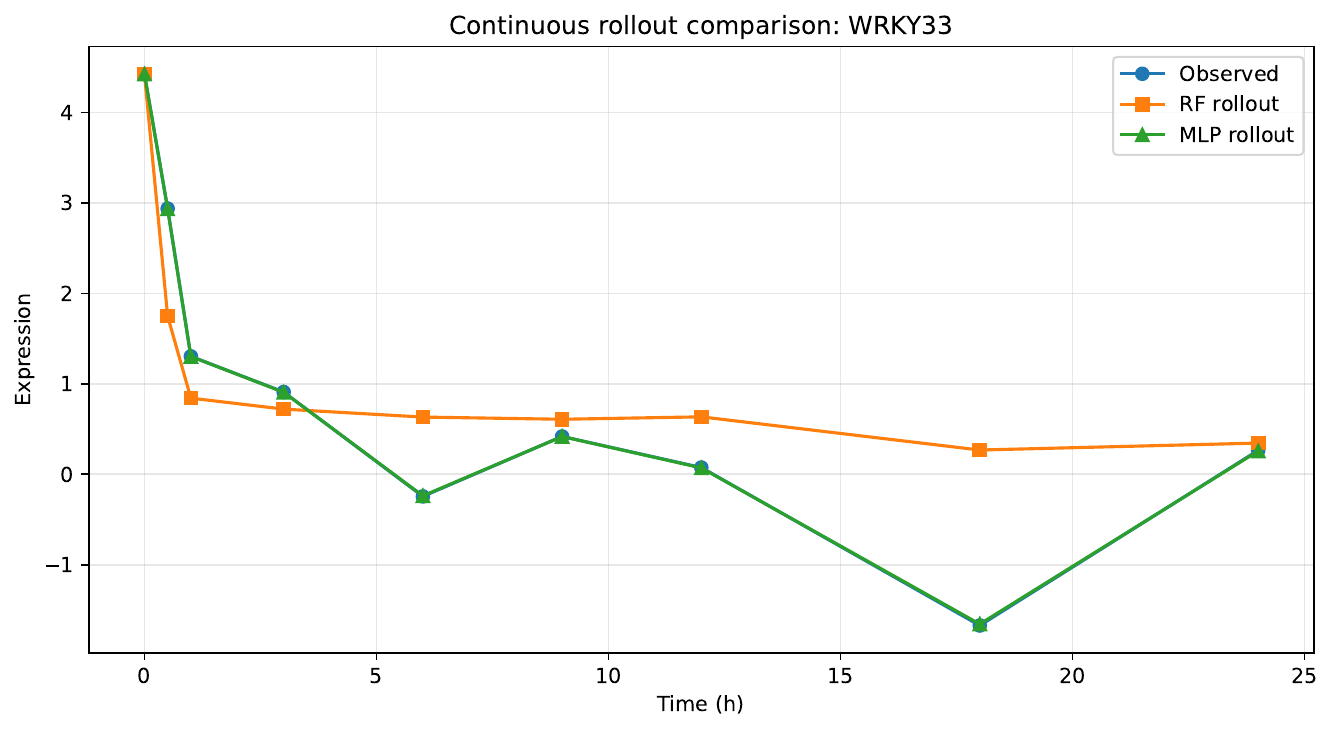}
    \includegraphics[scale=0.35]{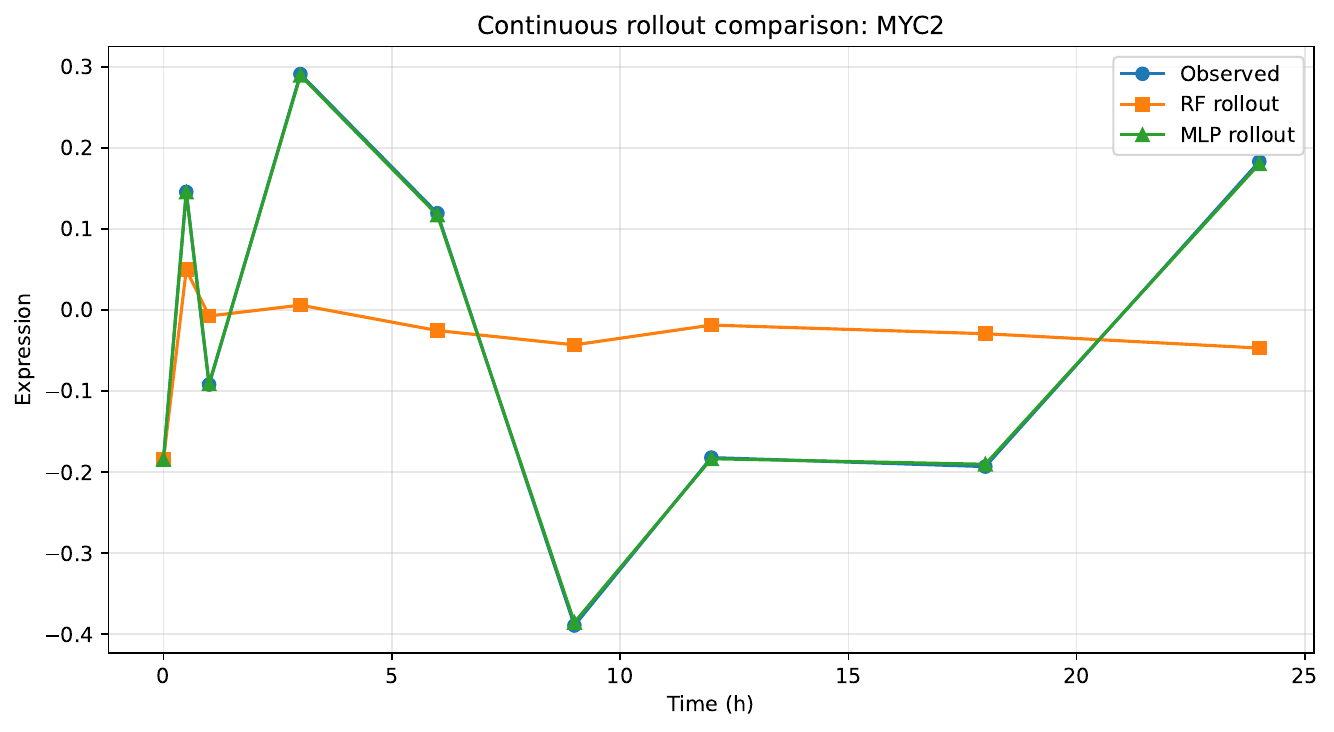}
    \includegraphics[scale=0.35]{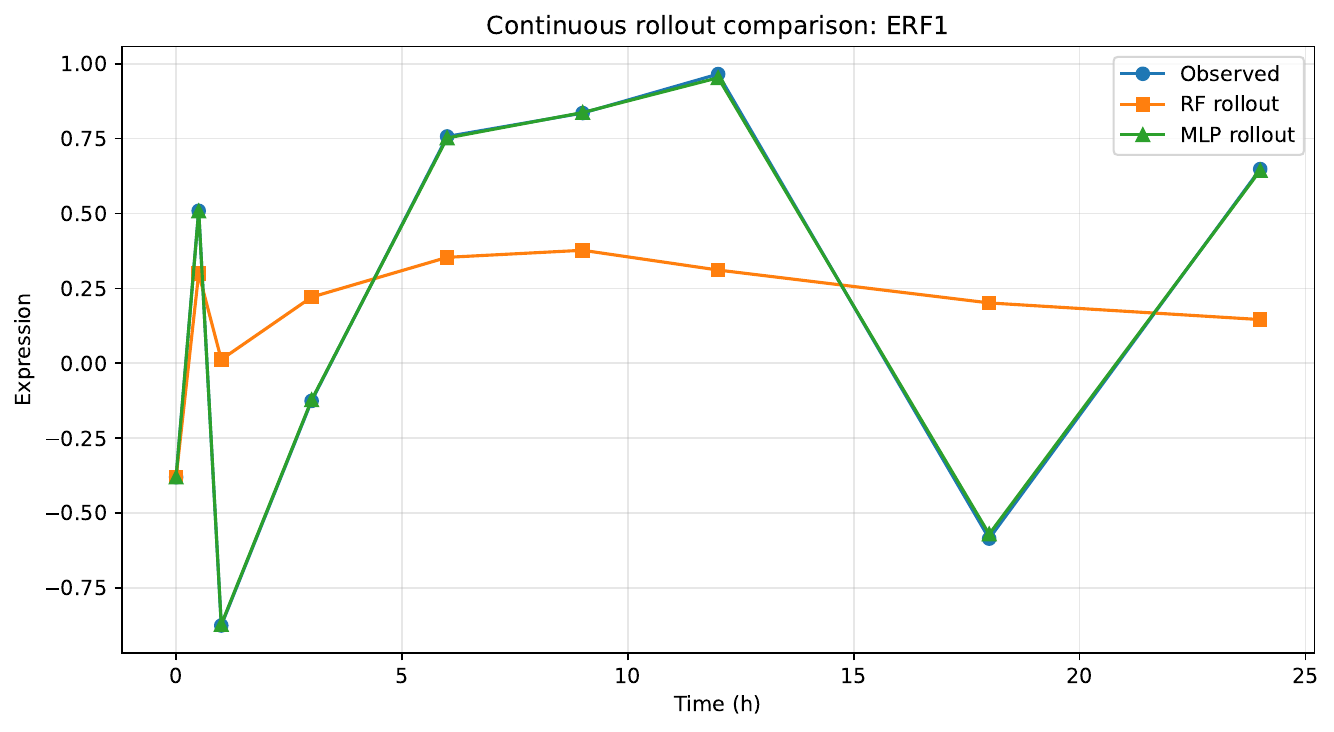}
    \includegraphics[scale=0.35]{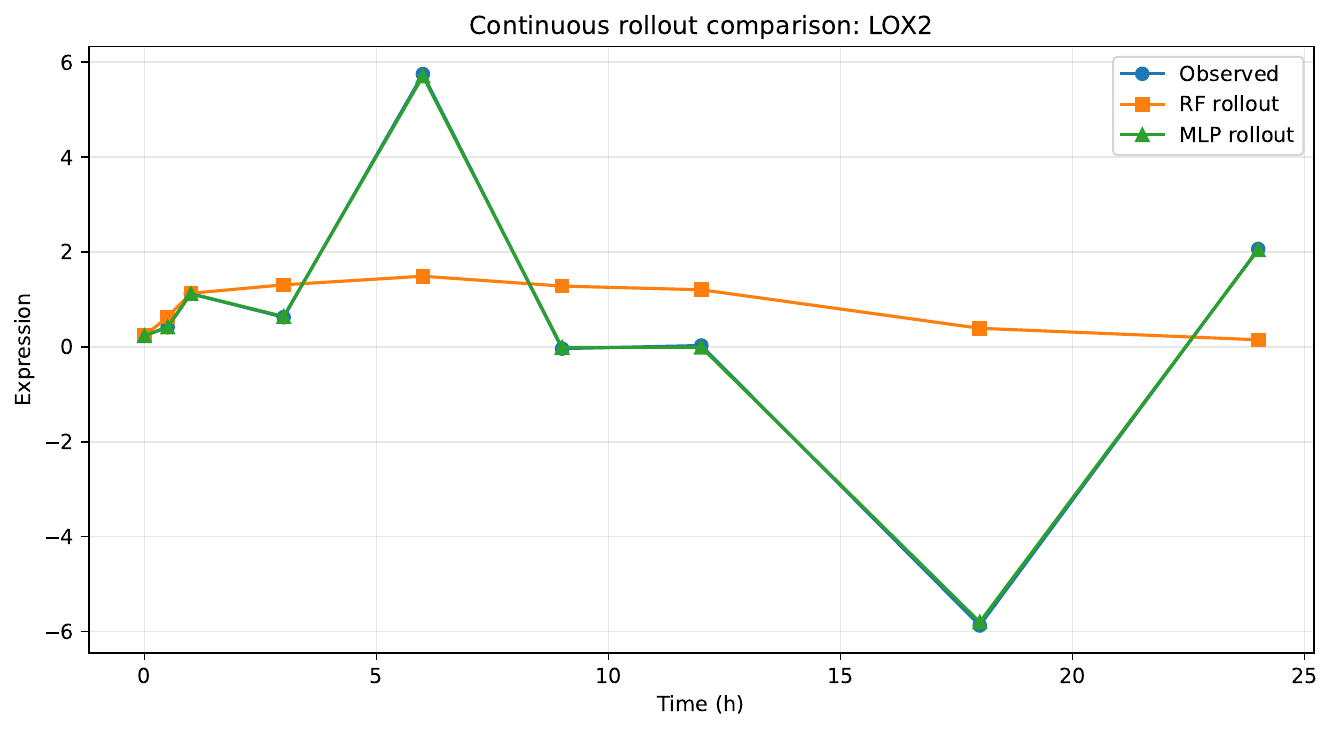}
    \caption{Continuous rollout comparison for the eight genes. Each panel shows the observed trajectory together with the recursive rollout generated by RF and MLP.}
    \label{fig:continuous_rollout}
\end{figure*}

\begin{figure*}[t]
    \centering
    \includegraphics[scale=0.35]{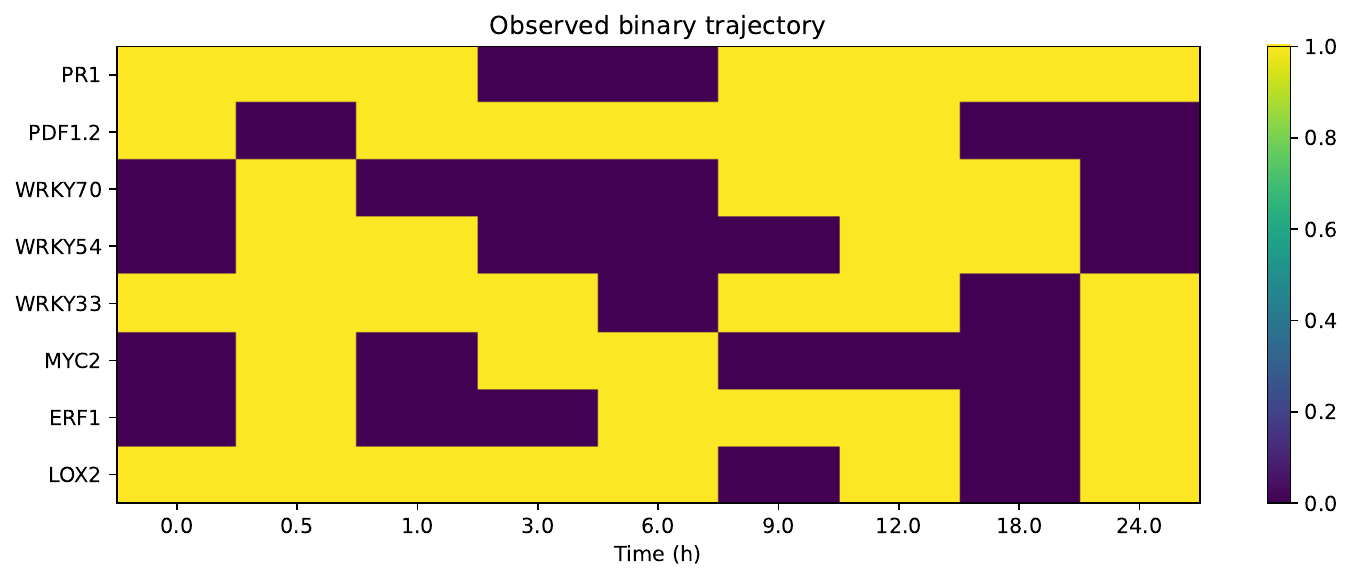}
    \includegraphics[scale=0.35]{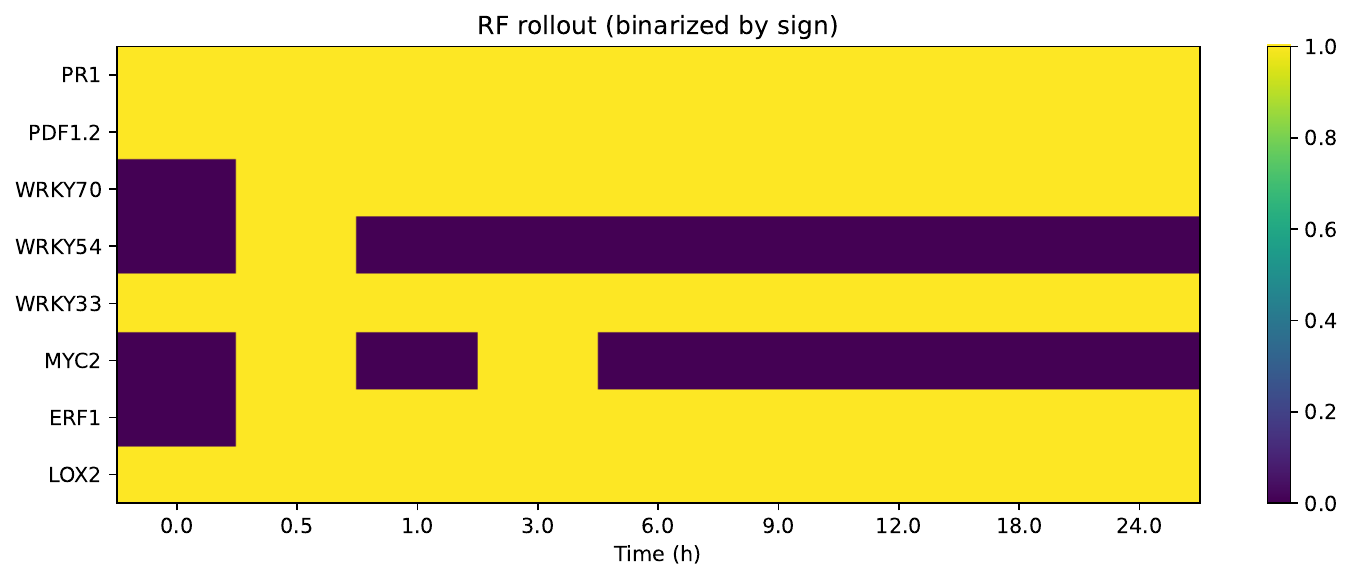}
    \includegraphics[scale=0.35]{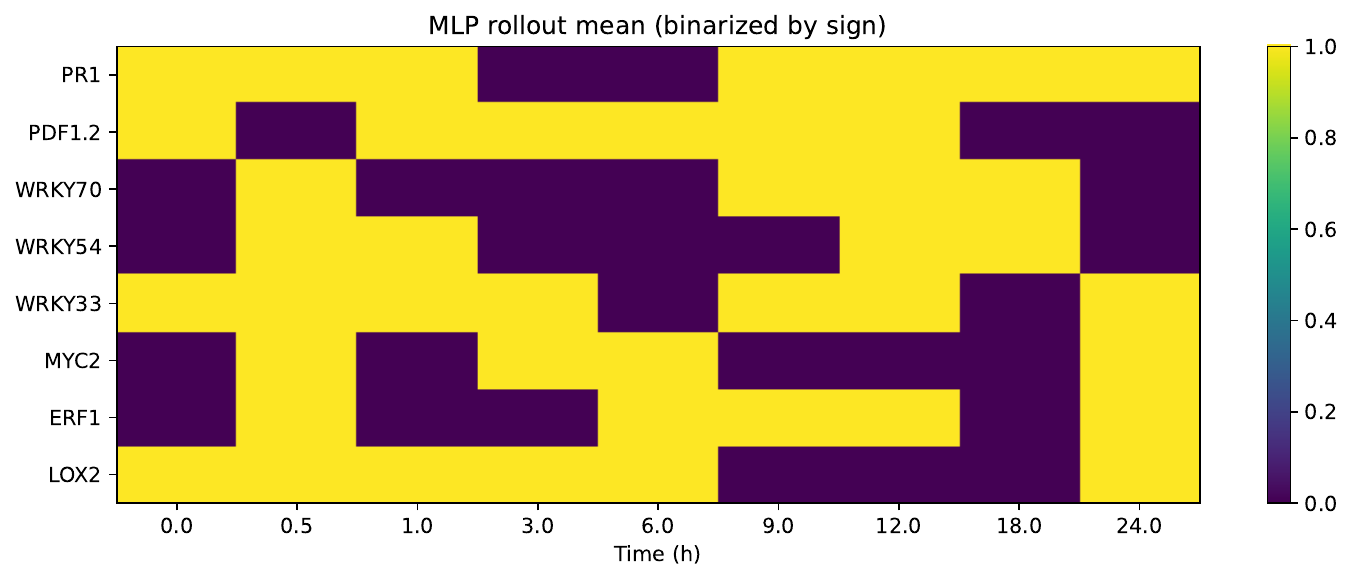}
    \includegraphics[scale=0.35]{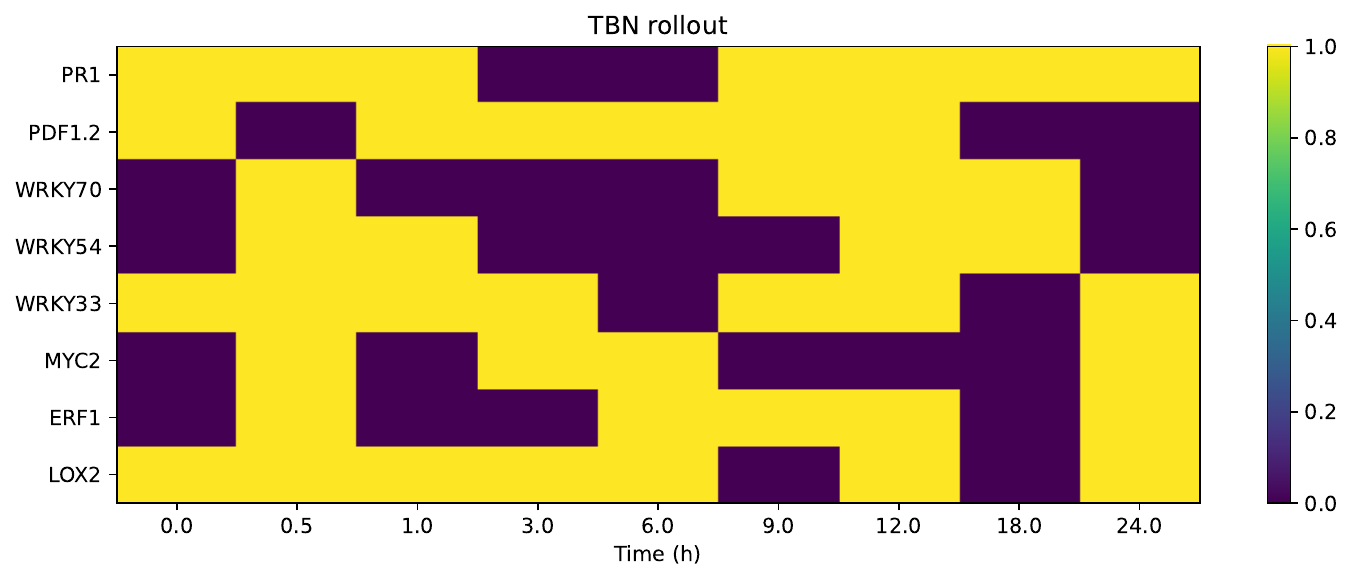}
    \caption{Observed and model-generated binary trajectories. From left to right: observed binarized trajectory, RF rollout binarized by sign, MLP rollout binarized by sign, and TBN rollout. The TBN reproduces the observed binary trajectory exactly, while the MLP remains very close and RF shows larger deviations.}
    \label{fig:binary_rollout_heatmaps}
\end{figure*}

The continuous rollout plots (Figure~\ref{fig:continuous_rollout}) show that the MLP tracked the observed temporal patterns substantially better than RF for most genes. This was particularly visible in genes whose trajectories exhibited stronger temporal variation. In turn, the binarized heatmaps (Figure~\ref{fig:binary_rollout_heatmaps}) provide a compact qualitative comparison: the TBN exactly reproduced the observed binary sequence, the MLP remained almost perfectly aligned with it, whereas RF introduced visible deviations over time.

\subsection{Interpretability analysis}

To provide an interpretable view of the learned models, we examined the RF feature importance visualization and the TBN weight matrix.

\begin{figure*}[t]
    \centering
    \includegraphics[scale=0.35]{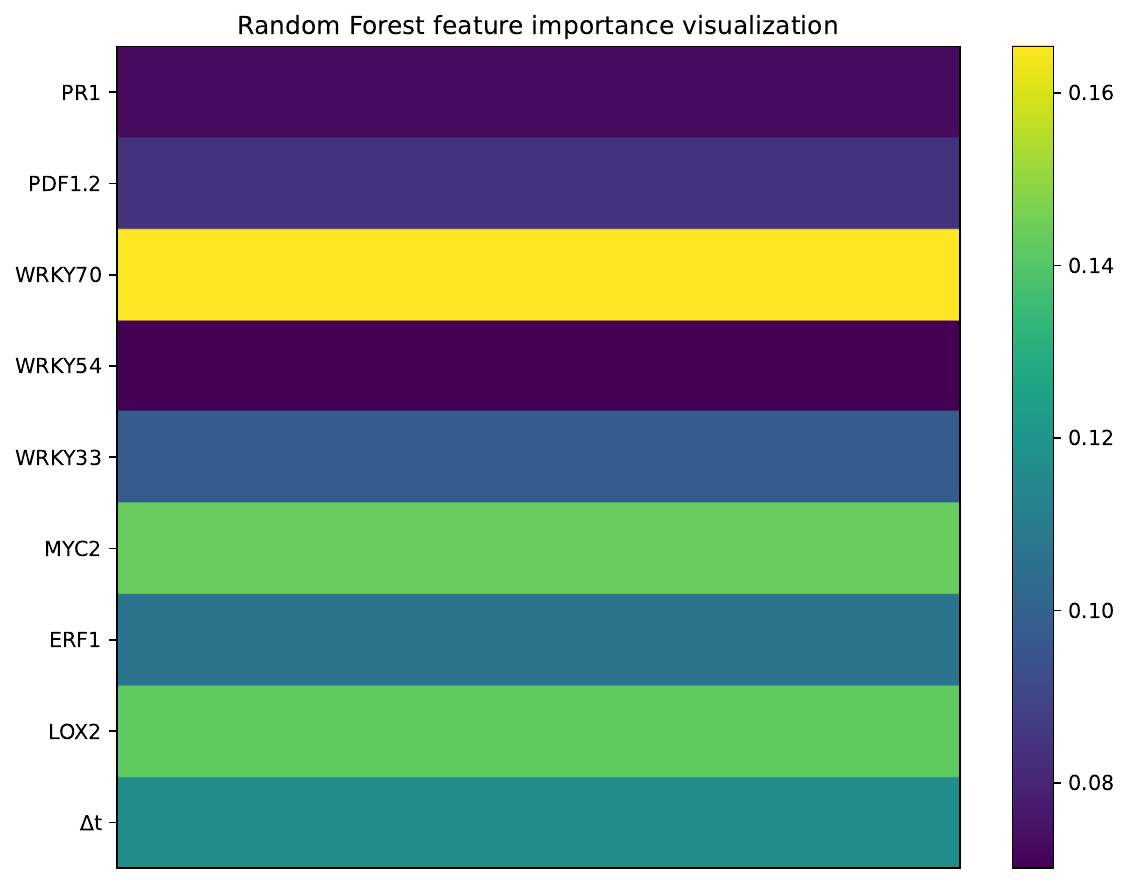}
    \includegraphics[scale=0.35]{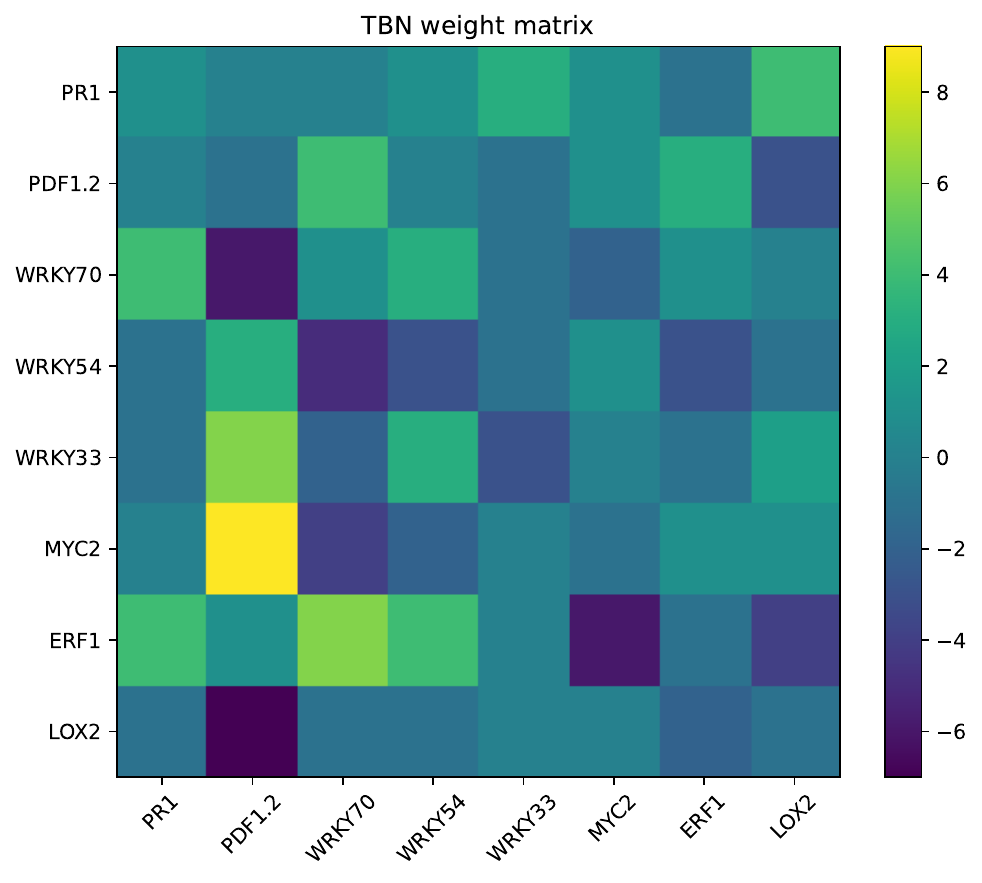}
    \caption{Interpretability analysis. Left: Random Forest feature importance visualization. Right: learned TBN weight matrix, where positive entries indicate putative activating interactions and negative entries indicate inhibitory interactions.}
    \label{fig:rf_importance_tbn_weights}
\end{figure*}

For RF, the importance analysis suggested that the contribution of the predictors was relatively diffuse, with no single dominant driver across outputs. Among the input variables, \textit{WRKY70}, \textit{MYC2}, \textit{LOX2}, and the time increment $\Delta t$ tended to have comparatively larger importance values than the remaining variables. Figure~\ref{fig:rf_importance_tbn_weights}(left) shows the RF feature importance matrix visualization.

For the TBN, interpretability is direct through the learned signed weight matrix. Positive values correspond to putative activating influences and negative values to inhibitory influences. The learned matrix (Figure~\ref{fig:rf_importance_tbn_weights}, right) shows a mixture of positive and negative interactions across the eight genes, indicating that the perceptron-based TBN was able to recover a nontrivial signed regulatory structure from the binarized transitions. In particular, some genes such as \textit{ERF1}, \textit{WRKY70}, \textit{WRKY33}, and \textit{MYC2} exhibited relatively large-magnitude incoming or outgoing weights, suggesting a stronger role in shaping the inferred discrete dynamics.

\subsection{Summary of empirical findings}

Overall, the experiments revealed three main findings. First, RF achieved the best average one-step numerical accuracy in the continuous domain. Second, the TBN achieved the best average one-step qualitative accuracy in the binary domain. Third, in recursive multi-step rollout, the TBN reproduced the observed binary trajectory exactly, while the MLP also showed near-perfect trajectory fidelity and RF accumulated substantially larger deviation. Taken together, these results show that the three modeling approaches behaved differently depending on whether the evaluation emphasized local numerical fit or global qualitative dynamical fidelity.

\section{Discussion}

This study compared two continuous surrogate models, RF and MLP, with a discrete mechanistic model based on a TBN on the same Arabidopsis ISR dataset. The main objective was not to identify a universally best predictor, but rather to examine how these modeling paradigms differ when evaluated from two complementary perspectives: local numerical prediction of gene-expression values and global qualitative reproduction of regulatory dynamics.

A first important observation is that the three models excelled under different criteria. RF achieved the best average one-step performance in the continuous domain, as reflected by the lowest MAE and RMSE. This suggests that, in this small-sample setting, the ensemble-based regressor was effective at capturing short-term nonlinear relationships in the raw expression values. However, when the evaluation was moved to the binary domain, RF no longer had a clear advantage. After sign binarization, its average one-step binary accuracy was similar to that of the MLP and lower than that of the TBN. This indicates that superior numerical fit in the continuous domain does not necessarily translate into better preservation of the qualitative regulatory state.

In contrast, the TBN achieved the best average one-step binary performance and, more importantly, reproduced the observed binarized trajectory exactly during recursive rollout. This is a notable result because the TBN was trained only on binarized transitions, not on the raw expression magnitudes. Its exact recovery of the binary trajectory suggests that the discrete mechanistic representation is particularly well suited to capturing the qualitative regulatory progression encoded by the sign structure of the data. From the perspective of Boolean network modeling, this is a desirable property, since the goal is not to reproduce exact expression amplitudes but to recover the sequence of active/inactive regulatory states and the corresponding interaction logic.

The MLP yielded perhaps the most interesting intermediate behavior. Although it did not outperform RF on average one-step numerical error, it produced a rollout trajectory that was almost perfectly aligned with the observed binary sequence, with only a single mismatch. This suggests that the MLP was able to learn a continuous state transition map whose recursive dynamics remained close to the sign structure of the biological process. In other words, the MLP did not have the strongest local numerical fit overall, but it generalized better than RF in the multi-step qualitative sense. This result highlights that one-step prediction metrics alone can be insufficient to characterize the usefulness of a model for dynamical biological systems.

Taken together, these findings support the view that continuous surrogates and discrete mechanistic models should be regarded as complementary rather than competing approaches. Continuous models such as RF and MLP are naturally suited to preserving amplitude information and may be preferable when the main goal is to predict expression levels directly. By contrast, TBNs provide an explicit representation of regulatory logic through signed weights and thresholds, and therefore are more directly interpretable in mechanistic terms. In the present case study, this mechanistic structure appears to be especially advantageous for preserving the qualitative temporal organization of the ISR response.

The interpretability analysis reinforces this distinction. The RF feature-importance visualization provided a useful indication of influential predictors, but the resulting importance patterns were relatively diffuse and do not directly encode regulatory sign or threshold effects. By comparison, the TBN weight matrix offers an immediate mechanistic interpretation: positive weights correspond to putative activating influences, negative weights to inhibitory influences, and the thresholds determine node activation conditions. This makes the TBN particularly attractive when the goal is not only prediction, but also formulation of regulatory hypotheses. This point is consistent with the broader use of threshold Boolean networks in gene regulatory network inference, where interpretability is often a central motivation \cite{10944722,Timmermann2020ISR}.

At the same time, the present results should be interpreted with appropriate caution. The dataset is extremely small, consisting of only nine time points and eight one-step transitions. Therefore, the conclusions should be understood as evidence from a proof-of-concept case study rather than as a definitive benchmark of model superiority. In particular, the strong rollout performance of the MLP may depend on the small scale and smoothness of this specific trajectory, while the exact trajectory reproduction by the TBN is facilitated by the sign-based binarization that abstracts away amplitude variability. More extensive validation on additional biological time-series datasets will be needed to determine how general these conclusions are.

A second limitation is that the continuous and discrete models are not trained on identical targets. RF and MLP are asked to predict real-valued expression vectors, whereas the TBN predicts only binarized next states. This asymmetry is intentional and reflects the distinct modeling philosophies under comparison, but it also implies that the models address slightly different notions of fidelity. For this reason, the most meaningful comparison is not based solely on MAE or RMSE, but on whether the inferred models preserve the biologically relevant qualitative dynamics. In this sense, the rollout analysis is especially informative.

Another important point is that the binarization rule itself imposes a strong abstraction on the data. In the ISR study, binarization was based on the sign of the normalized expression values, mapping positive values to activation and negative values to repression \cite{Timmermann2020ISR}. This representation is biologically appealing because it captures direction of regulation, but it inevitably discards information about expression magnitude. Consequently, exact agreement in the binary domain should not be interpreted as full reconstruction of the continuous biological process. Rather, it indicates accurate recovery of the qualitative regulatory pattern.

From a broader perspective, the main contribution of this study is methodological. The results show that it is possible to compare continuous and discrete models within a unified experimental setting by using the same underlying biological dataset in two representations: raw continuous expression and sign-binarized states. This type of comparison makes explicit the tradeoff between predictive fidelity and mechanistic interpretability that is often implicit in computational biology. In the present case, RF was strongest in local continuous prediction, TBN was strongest in discrete qualitative fidelity, and MLP occupied an intermediate position with especially strong multi-step behavior. This three-way contrast would have been missed if the evaluation had been restricted to only one type of metric.

Finally, these findings suggest several directions for future work. One natural extension is to evaluate the same comparison on additional gene regulatory datasets, ideally including larger systems and longer time series. A second direction is to investigate hybrid strategies, for example using continuous models to capture expression amplitudes while extracting qualitative regulatory logic from their binarized or thresholded dynamics. A third direction is to strengthen the TBN side by moving beyond a perceptron-based baseline toward inference methods that explicitly incorporate attractor constraints or sparsity criteria. Such extensions could help clarify under what conditions continuous surrogates and discrete mechanistic models converge to similar dynamical conclusions, and when they instead provide fundamentally different views of the same biological process.

\section{Conclusion}

In this work, we compared continuous surrogate models and a discrete mechanistic model on the same Arabidopsis ISR gene-expression dataset, using both the raw continuous measurements and their sign-binarized representation. The results showed that Random Forest achieved the best average one-step numerical performance in the continuous domain, whereas the threshold Boolean network obtained the best average one-step qualitative performance in the binary domain and reproduced the observed binarized trajectory exactly during recursive rollout. The Multi-Layer Perceptron showed an intermediate but noteworthy behavior, yielding near-perfect rollout fidelity despite not having the best average one-step numerical scores.

Overall, these findings suggest that continuous and discrete models provide complementary views of the same biological process. Continuous surrogates are useful for capturing expression magnitudes, while threshold Boolean networks offer a more interpretable representation of qualitative regulatory dynamics. This proof-of-concept study supports the value of evaluating both predictive accuracy and dynamical fidelity when modeling gene regulatory systems, and motivates future work on larger datasets and hybrid modeling strategies that combine the strengths of both approaches.

\section*{Acknowledgment}
The author acknowledges the use of ChatGPT for grammar checking and coding assistance in the preparation of this manuscript. The scientific content, analysis, and conclusions remain the sole responsibility of the author.

\section*{Data and Code Availability}
The data and code used in this study are publicly available at the following GitHub repository: \url{https://github.com/gruzh/CS_vs_TBN_ISR/}.

\bibliographystyle{IEEEtran}
\bibliography{bibliography_CIBCB2026_file}

\end{document}